\input{aipcheck}

\documentclass[
final            
  ]
  {aipproc}

\layoutstyle{6x9}

\begin{document}

\title{Doorway States and Billiards}


\author{J.~A.~Franco-Villafa\~ne}{
  address={Instituto de Ciencias F\'isicas, Universidad Nacional
Aut\'onoma de M\'exico, P.O. Box 48-3, 62251 Cuernavaca Mor., Mexico.}
}

\author{J.~Flores}{
  address={Instituto de F\'isica, Universidad Nacional Aut\'onoma de M\'exico, P.O. Box 20-364, 01000 M\'exico, D. F., Mexico.}
	,altaddress={Centro Internacional de Ciencias, A. C., P.O. Box 6-101 C.P. 62131 Cuernavaca, Mor. Mexico.}
}

\author{J.~L.~Mateos}{
  address={Instituto de F\'isica, Universidad Nacional Aut\'onoma de M\'exico, P.O. Box 20-364, 01000 M\'exico, D. F., Mexico.}
}

\author{R.~A.~M\'endez-S\'anchez}{
  address={Instituto de Ciencias F\'isicas, Universidad Nacional
Aut\'onoma de M\'exico, P.O. Box 48-3, 62251 Cuernavaca Mor., Mexico.}
}

\author{O.~Novaro}{
  address={Instituto de F\'isica, Universidad Nacional Aut\'onoma de M\'exico, P.O. Box 20-364, 01000 M\'exico, D. F., Mexico.},altaddress={Member of El Colegio Nacional}
}

\author{T.~H.~Seligman}{
  address={Instituto de Ciencias F\'isicas, Universidad Nacional
Aut\'onoma de M\'exico, P.O. Box 48-3, 62251 Cuernavaca Mor., Mexico.}
	,altaddress={Centro Internacional de Ciencias, A. C., P.O. Box 6-101 C.P. 62131 Cuernavaca, Mor., Mexico.}
}

\begin{abstract}
Whenever a distinct state is immersed in a sea of complicated and dense states, the strength of the distinct state, which we refer to as a doorway, is distributed in their neighboring states. We analyze this mechanism for 2-D billiards with different geometries.  One of them is symmetric and integrable, another is symmetric but chaotic, and the third has a capricious form. The fact that the doorway-state mechanism is valid for such highly diverse cases, proves that it is robust.
\end{abstract}

\maketitle


{\em \noindent We dedicate this paper to the memory of Marcos Moshinsky.}

\section{INTRODUCTION}

\begin{figure}[h!]
  \includegraphics[width=0.8\columnwidth]{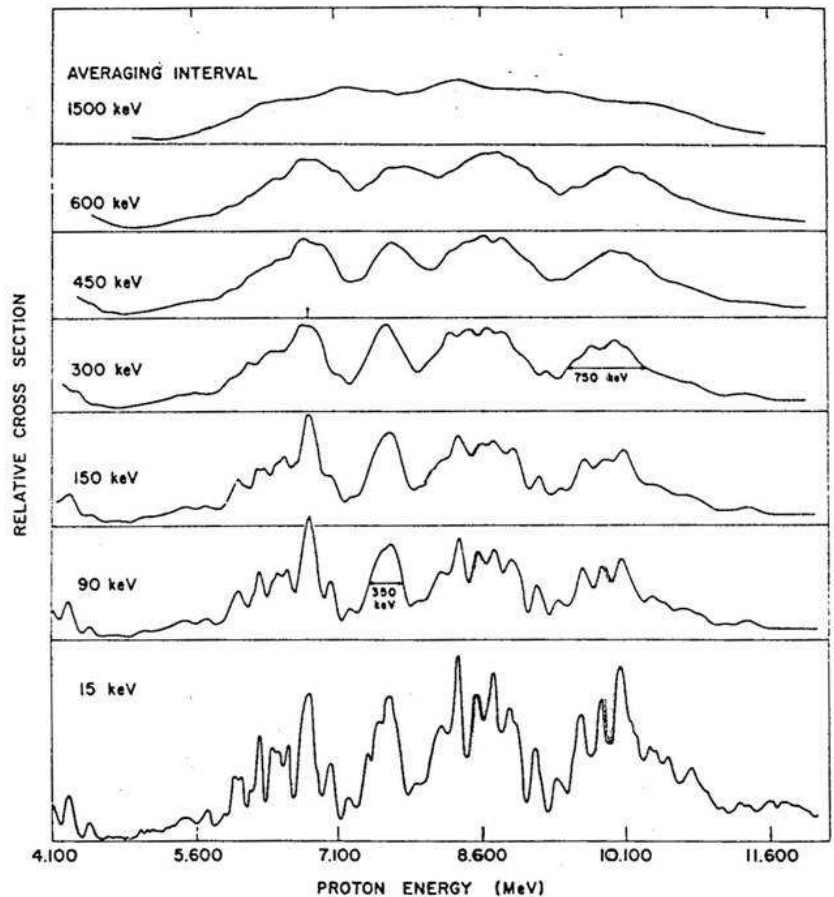}
  \caption{A very well known nuclear physics example of a giant resonance, intermediate and fine structure obtained with the nuclear reaction $^{27}\mathrm{Al}(\mathrm{p},\gamma)^{28}\mathrm{Si}$. As the average interval diminishes the fine structures appears.}\label{fig1}
\end{figure}

The doorway state mechanism, which was introduced in nuclear physics a long time ago~\cite{Brown1964}, is effective when a ``distinct'' state is coupled to the scattering channel and also to a sea of more dense and complicated states. Prime examples of doorway states are isobaric analogue states and giant resonances in nuclei. The results given in Fig.~\ref{fig1} for the photonuclear reaction $^{27}\mathrm{Al}(\mathrm{p},\gamma)^{28}\mathrm{Si}$ are well known for nuclear physicists and show that the cross section acquires more structure as the energy average decreases. In the last few years, this mechanism has been found in many instances. Doorway states have been observed in many quantum systems: atoms and molecules~\cite{Kawata2000}, clusters~\cite{Hussein2000}, quantum dots~\cite{Baksmaty2008}, and in C60 fullerenes \cite{Laarmann2007,Hertel2009}. Furthermore, they also appear in classical wave systems: flat microwave cavities with a thin barrier inside~\cite{Aberg2008}, and even also in the seismic response of sedimentary valleys~\cite{Flores}.

The doorway state is not an eigenstate of the whole system. Therefore, the strength of the doorway state spreads among the eigenstates within some energy region around the distinct state energy. This produces a  strength function whose width is commonly known as {\em spreading width}. In the simplest case, the strength function follows the Breit-Wigner form, a Lorentzian, as a function of energy~\cite{Bohr1969}. 

We therefore see that the doorway state concept is a unifying one in physics covering a wide range of quantum and classical systems, established on scales ranging from fermis to tens of kilometers. In this paper we shall analyze this concept from another point of view, discussing its occurrence in two-dimensional billiards of different shapes: a rectangle, a stadium, and one that resembles the irregular shape of a sedimentary basin~\cite{Flores1987}. The first is symmetric and integrable, the second is symmetric but chaotic, and the last one shows neither integrability nor symmetry. We will show numerically that the doorway state mechanism is indeed robust.

\section{II. A MODEL FOR THE ELASTIC DOORWAY STATE}

In order that a system shows the phenomena described by the strength function, the presence of a doorway state and a sea of complicate states must be assured. In the classical systems analyzed up to now, this has been achieved in different ways. In the case of microwave cavities, a doorway has been produced by introducing a thin barrier inside a rectangular cavity. The barrier produces several states, called superscars, which have been detected experimentally and act as doorway states~\cite{Aberg2008}. A different system is a rigid parallelepiped cavity with an elastic membrane on one of its walls. This is an acoustic realization of the doorway states, provided now by the normal-mode states of the elastic membrane coupled to the denser states of the fluid inside the otherwise rigid walls forming the cavity. In this case~\cite{Flores1992}, the membrane states act literally as doorways, since the energy of the scattering channel enters the fluid inside the cavity when the membrane excites the interior acoustic states. An elastic doorway state is also present when the seismic response of sedimentary basins covered by a bounded region of soft material is considered, as we shall now describe. The analysis of this response will serve us as a guide to introduce in billiards a model to study the strength-function.

What is the doorway state in the seismic case? It has been known for a long time to geophysicists~\cite{Ewing1957} that when a sedimentary layer is covered by a much softer material (such as happens, for example, on the ocean floor) a coupling occurs between evanescent SP waves in the soft layer and Rayleigh-type waves on the interface. The coupling condition is~\cite{Lomnitz1999}
\begin{equation}
 0.91\beta_{1}<\alpha_{0}=v,
\end{equation}
where $v$ is the phase velocity of the coupled mode and $\beta_{1}$, $\alpha_{0}$ are the S-wave velocity in the sediment and the P-wave velocity in the softer region, respectively. The coupling occurs when the phase velocity of the dispersive Rayleigh waves is equal to the sound velocity $\alpha_{0}$ of the very soft terrains. The coupled mode has many features in common with what is known as an Airy phase~\cite{Ewing1957}. In particular, they are monochromatic and of long duration. We have called this mode a PR mode~\cite{Lomnitz1999}.

If the soft-clay terrain is bounded, as happens for example in the Mexico City basin as well as in San Francisco, Kobe and many other cities around the world, once established, the PR mode reflects at the soft clay boundaries due to the large impedance contrast between the clays and the sediments surrounding them, and the very fact that the boundary layer on and near where it lives, terminates at this boundary. These surface waves have to be evanescent outside the interface, they imply horizontal compression movement in the soft clay above the interface, and thus strongly couple to several normal modes in the soft clay bed located in the horizontal $xy$-plane. These modes provide the sea of complicated states~\cite{Mateos1993}.

Since the source of the seismic waves is far away from the basin, we represent the PR mode by a plane wave $\exp(i\mathbf{k}\cdot\mathbf{r})$, where the direction of $\mathbf{k}$ varies for different earthquakes corresponding to different epicenters. Here $\mathbf{r}$ fixes a point within the bounded soft terrain in the horizontal $xy$-plane. The magnitude of $\mathbf{k}$ for the PR mode is $k=2\pi\nu/\alpha_{0}$.

What is the sea of dense and complicated states in the seismic case? The sea is formed by the normal-mode states with amplitudes $\phi_{i}$ and frequencies $\nu_{i}$ that correspond to the eigenfunctions of a 2-D Helmholtz equation
\begin{equation}
 \nabla^{2}\phi_{i}+k_{i}^{2}\phi_{i}=0
\end{equation}
in the region of the soft region. Here $k_{i}=2\pi\nu_{i}/\alpha_{0}$, where $\alpha_{0}$ is the P-wave velocity in the clays. We use Neumann boundary conditions $\hat{n}\cdot\nabla\phi_{i}=0$, where $\hat{n}$ is a vector normal to the boundary. Note that the Neumann conditions are somewhat arbitrary, but a similar calculation with Dirichlet conditions yields qualitatively the same results.


The spreading of the doorway state $\exp(i\mathbf{k}\cdot\mathbf{r})$ among the normal modes $\phi_{i}$ is then given by
\begin{equation}
 A(\nu_{i})=\left|\int\exp(i\mathbf{k}\cdot\mathbf{r})\phi_{i}(x,y)dxdy\right|^{2}
\end{equation}
In a few words, our model for the doorway states in two-dimensional billiards is the following: a monochromatic plane wave $\exp(i\mathbf{k}\cdot\mathbf{r})$, with $k=2\pi\nu/\alpha_{0}$ represents the doorway state and the sea of complicated states is formed by the eigenstates $\phi_{i}$ of a billiard with a given boundary, phase velocity $\alpha_{0}$ and Newmann boundary conditions.

\section{III. SPREADING WIDTH IN BILLIARDS}

\begin{figure}[h!]
  \includegraphics[width=0.8\columnwidth]{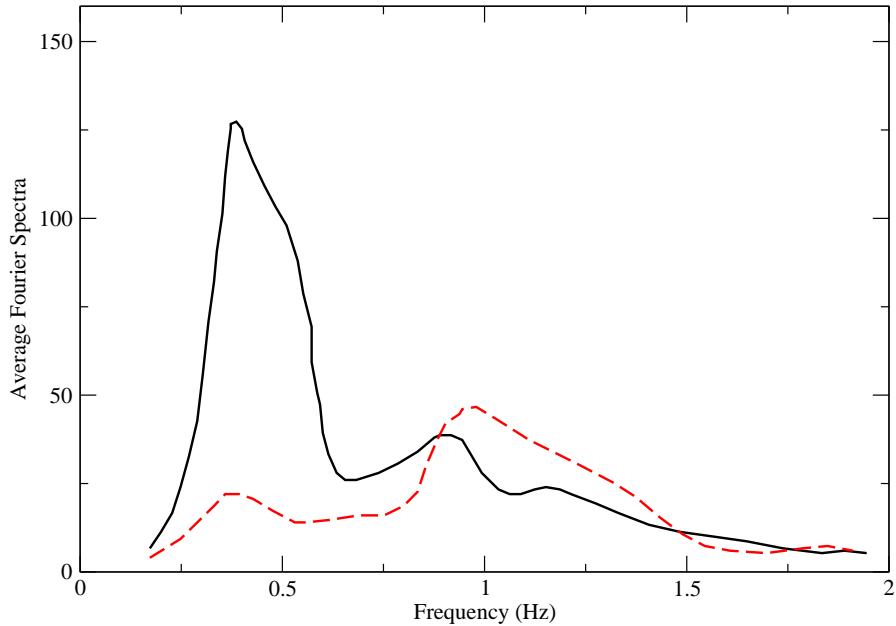}
  \caption{Averaged Fourier spectrum obtained from seismograms of the earthquake measured in Mexico City of magnitude 6.9, 1989 April 25. The continuous line corresponds to an average of stations located in the old lake bed; the dashed line to stations located in the rock zone.}
\label{fig2}
\end{figure}

\begin{figure}[h!]
  \includegraphics[width=0.8\columnwidth]{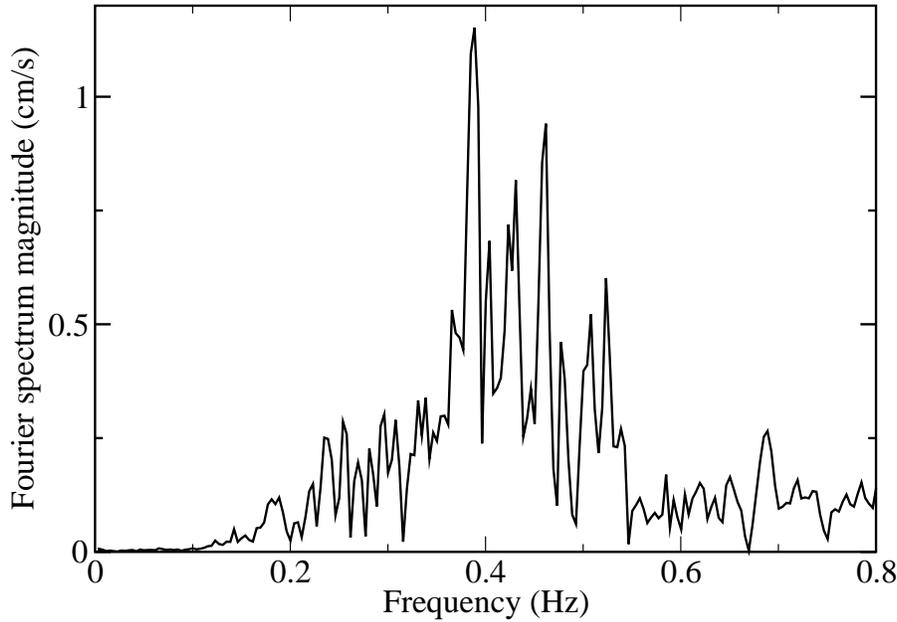}
  \caption{Fourier spectrum obtained from a seismogram measured in a station located on the lake bed clay in Mexico City during the earthquake of magnitude 7.1, 1997 January 11.}
\label{fig3}
\end{figure}

In order to determine the parameters entering the calculations, we shall use data corresponding to the seismic response of the Valley of Mexico. This is a basin formed by hard enough sedimentary deposits ($\alpha_{1}=3000$m/s, $\beta_{1}=1500$m/s) on top of which there is a very soft terrain ($\alpha_{0}=1500$m/s, $\beta_{0}=50$m/s) which was formerly a lakebed. The shape of the Tenochtitlan lake, where Mexico City lies, is well established. The seismic response of this basin will be seen in Figs.~\ref{fig2} and~\ref{fig3}, for low and high resolution. In any case, it is seen that $\nu=0.4$Hz is observed in all strong earthquakes in Mexico City. One should note the analogies between Figs.~\ref{fig1} and~\ref{fig2},~\ref{fig3}, which at some level justifies the assumption that a strength function appears.

We now present the resulting strength function $A(\nu_{i})$ for three different billiards: a rectangle, which is symmetric and integrable; a stadium, which has much the same symmetry as the rectangle, but which is chaotic; and a third one, which is neither symmetric nor integrable and has the shape of the lake. The linear dimensions of the three systems are comparable and $\alpha_{0}$, $\nu$, which depend on the sediment or clay properties, are taken to be equal to the appropriate values for the lake.

The strength function for the first case can be obtained analytically and for the two other cases it was obtained numerically, using a finite element method to obtain the wave amplitudes $\phi_{i}$. The numerical results were calculated using the finite element method with linear polynomials. The region was discretized using 8272 points located in a rectangular grid. The size of the grid is 118 m. We verified the accuracy of the program to be better than one percent using a region with a rectangular boundary, which can be computed analytically.

\begin{figure}[h!]
  \includegraphics[width=0.8\columnwidth]{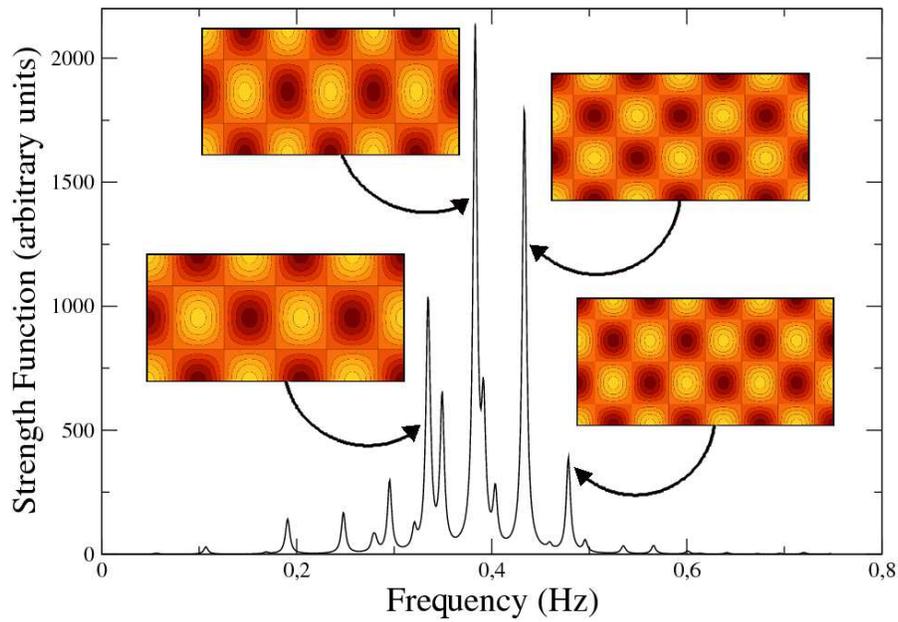}
  \caption{Strength function phenomenon for a rectangular billiard.}\label{fig4}
\end{figure}
\begin{figure}[h!]
  \includegraphics[width=0.8\columnwidth]{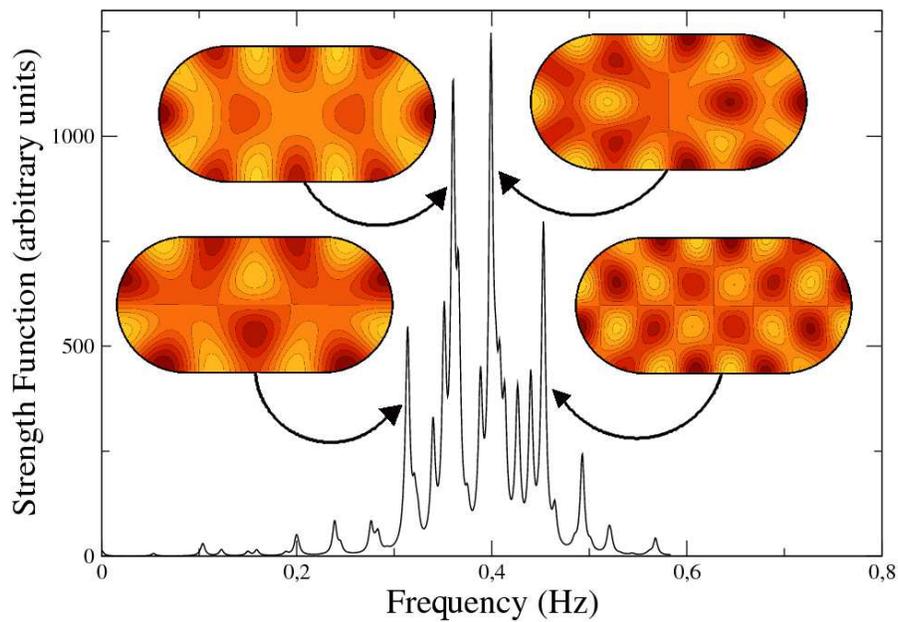}
  \caption{Strength function phenomenon for a stadium.}\label{fig5}
\end{figure}
\begin{figure}[h!]
  \includegraphics[width=0.8\columnwidth]{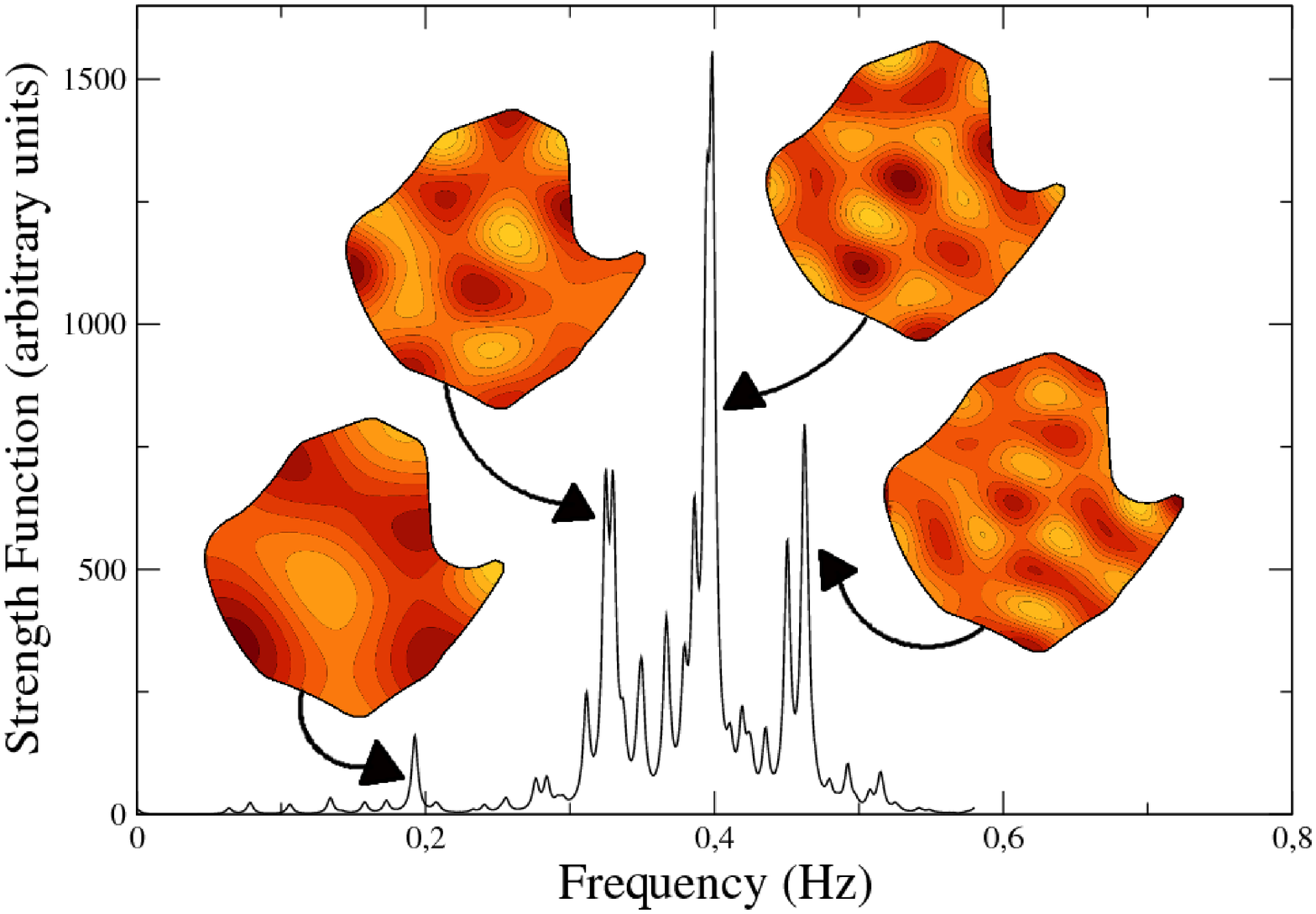}
  \caption{Strength function. Some of the wave amplitudes of the lake, that correspond to the peaks of $A$, are shown.}\label{fig6}
\end{figure}

In Figs.~\ref{fig4} to~\ref{fig6} we show that a strength function $A$ appears for the three systems considered, it is plotted as a function of frequency. The peaks in this figure are related  to the wave amplitudes shown as insets. In particular, we see from Fig.~\ref{fig3} that our calculation agrees qualitatively with what is observed in many seisms in Mexico City. In any case the strength function appears independently of the billiard symmetry, its integrability or lack of. It can therefore be deemed robust.

\section{Conclusions}

As a conclusion, one can say doorway states appear in very diverse systems and circumstances. They show up in many quantum and classical systems and also in billiards of quite different shapes.  Their existence has equivalent consequences on the response of the system not regarding whether this is integrable or chaotic, symmetric or arbitrary. Furthermore, they also appear covering many  orders of magnitude in the characteristic length of the system. As a matter of fact, the results of the seismic response of sedimentary valleys imply that these lengths range from fermis up to tens of kilometers.

\begin{theacknowledgments}
  This work was supported by DGAPA-UNAM under project PAPIIT-IN111308 and by CONACyT under project 79613.
\end{theacknowledgments}

\bibliographystyle{aipproc}   

\end{document}